\pdfoutput=1
\documentclass[a4paper,fleqn]{cas-dc}

\usepackage{cite}
\usepackage{subfigure}
\usepackage[ruled,vlined,algo2e]{algorithm2e}
\usepackage{algorithm}
\usepackage{algorithmic}

\usepackage[numbers, sort]{natbib}

\def\tsc#1{\csdef{#1}{\textsc{\lowercase{#1}}\xspace}}
\tsc{WGM}
\tsc{QE}

\begin{document}
\let\WriteBookmarks\relax
\def\floatpagepagefraction{1}
\def\textpagefraction{.001}





\title [mode = title]{Semantic Communication for Cooperative Perception based on Importance Map}  



%

\author[1]{Yucheng Sheng}
\ead{shengyucheng@seu.edu.cn}
\author[2]{Hao Ye}
\ead{yehao@ucsc.edu}
\author[1,3]{Le Liang}
\cormark[1]
\ead{lliang@seu.edu.cn}
\author[1]{Shi Jin}
\cormark[1]
\ead{jinshi@seu.edu.cn}
\author[4]{Geoffrey Ye Li}
\ead{geoffrey.li@imperial.ac.uk}

\address[1]{National Mobile Communications Research Laboratory, Southeast University, Nanjing 210096, China}

\address[2]{Department of Electrical and Computer Engineering, University of California, Santa Cruz, CA 95064}

\address[3]{Purple Mountain Laboratories 211100, Nanjing, China}

\address[4]{ITPLab, the Department of Electrical and Electronic Engineering, Imperial, College London, London, U.K.}
\cortext[1]{Corresponding author}
\begin{abstract}
Cooperative perception, which has a broader perception field than single-vehicle perception, has played an increasingly important role in autonomous driving to conduct 3D object detection. Through vehicle-to-vehicle (V2V) communication technology, various connected automated vehicles (CAVs) can share their sensory information (LiDAR point clouds) for cooperative perception. We employ an importance map to extract significant semantic information and propose a novel cooperative perception semantic communication scheme with intermediate fusion. Meanwhile, our proposed architecture can be extended to the challenging time-varying multipath fading channel. To alleviate the distortion caused by the time-varying multipath fading, we adopt explicit orthogonal frequency-division multiplexing (OFDM) blocks combined with channel estimation and channel equalization. Simulation results demonstrate that our proposed model outperforms the traditional separate source-channel coding over various channel models. Moreover, a robustness study indicates that only part of semantic information is key to cooperative perception. Although our proposed model has only been trained over one specific channel, it has the ability to learn robust coded representations of semantic information that remain resilient to various channel models, demonstrating its generality and robustness.
\end{abstract}




\maketitle


\section{Introduction}
Dating back to the 1940s, Shannon's separation theorem demonstrates that the separate source and channel coding can reach optimality with an infinite block length \cite{shannon}. However, an increasing number of wireless applications, such as Internet-of-Things and autonomous driving, require low latency real-time communication and low computation complexity, where the infinite coding length is not practical. Semantic communication with joint source-channel coding (JSCC) is proposed to optimize source coding and channel coding to achieve better performance for specific tasks. Inspired by the significant success of the deep learning techniques, JSCC is first introduced to text transmission through recurrent neural networks (RNN) \cite{farsad2018deep}. In \cite{jiang2022deep}, hybrid automatic repeat request (HARQ) is exploited to reduce semantic transmission error further. In addition to the reconstruction task, a multi-task semantic communication system is proposed for various text tasks in \cite{multi-task}. Meanwhile, for wireless image transmission, a deep JSCC system \cite{jscc-image} is proposed, which directly maps the image pixel values to the complex-valued channel input symbols instead of relying on explicit coding for either compression or error correction. Later, the proposed deep JSCC system is extended to feedback channels \cite{DeepJSCC-f}, transmission with digital constellation \cite{DeepJSCC-Q}, or image retrieval problem at the edge \cite{retrieval}. Similar to image transmission, the JSCC schemes \cite{DeepWiVe}, \cite{Video} for video transmission map video signals to channel symbols, combining video compression, channel coding, and modulation steps into a single neural transform.

In addition to text, speech \cite{han2022semanticspeech}, images \cite{jscc-image} and videos \cite{DeepWiVe}, sensor data (LiDAR point clouds) widely in autonomous driving also contains essential semantic information \cite{xu2022opv2v}, \cite{xu2022cobevt}. Since the perception field of the autonomous car itself is usually limited by the objects, such as buildings and trees, a novel detection scheme with cooperative perception is investigated for a broader perception field. Through the utilization of V2V communication technology, different connected automated vehicles (CAVs) can exchange and fuse their sensory information, enabling the provision of multiple viewpoints for the same obstacle to complement one another \cite{xu2022opv2v}, \cite{xu2022cobevt}, \cite{zhou2022multi}, \cite{arnold2020cooperative}, \cite{tu2022maxim}, \cite{li2023learning}. The exchanged information encompasses raw data, intermediate features, detection outputs from individual CAVs, and metadata such as timestamps and poses. From the perspective of the fusion method, cooperative perception can be categorized into early fusion, intermediate fusion, and late fusion. Early fusion of raw data (LiDAR point clouds) requires the most communication resources. Intermediate fusion of the intermediate feature extracted from the raw data requires fewer communication resources than early fusion but can potentially perform as well as early fusion. Late fusion transmits the detection outputs from the CAV, which requires the least amount of resources at the expense of degraded performance. The intermediate fusion is usually considered in cooperative perception. Since the data size of shared information in cooperative perception and the number of CAVs can be huge and surpass the communication capability of conventional vehicular links, it is urgent to optimize the trade-off between perception performance and communication overhead.

To balance performance and communication bandwidth, several works have put forth solutions based on intermediate fusion from different perspectives. A handshake mechanism that selects the most relevant CAVs is proposed in \cite{liu2020when2com}, which learns how to construct communication groups and decide when to communicate. Unlike broadcast-based methods, the approach is inspired by general attention mechanisms and involves decoupling the communication stages, leading to a reduction in the amount of transmitted data. Meanwhile, an end-to-end learning-based source coding method is considered in \cite{xu2022v2x}, which utilizes a spatially aware graph neural network (GNN) to aggregate the information received from the CAVs. A 1D convolution is adopted in \cite{li2021learning} to compress the message, which is then pushed to the communication channels. Nevertheless, all previous studies have made the assumption that once two agents collaborate, they are required to share perceptual information of all spatial areas equally. However, this may significantly waste bandwidth, as a considerable portion of spatial areas may contain irrelevant information for the perception task. Hence, a spatial-confidence-aware communication strategy is proposed in \cite{hu2022where2comm} to transmit the relevant semantic information, which can reduce the communication bandwidth substantially.

These works typically assume perfect communications between CAVs and ignore underlying channel effects. A new late fusion method \cite{liu2023cooperative} transmits intermediate features over the wireless channel to fuse the detection outputs. The communication overhead of the new late fusion method is as large as the intermediate fusion method. However, the perception accuracy is less than that of the intermediate fusion because the essence of the new late fusion is a kind of late fusion. Hence, it is urgent to propose a novel intermediate fusion method over the wireless channel to overcome the distortion caused by the channel impairments, reduce the communication overhead, and achieve high perception accuracy.

 In response, we introduce a cooperative perception semantic communication framework, which employs an importance map to extract significant semantic information at the transmitter and fuses the intermediate feature through an attention-based mechanism at the receiver. Our proposed system is designed based on a JSCC architecture and trained in an end-to-end learning manner, which is optimized to achieve better semantic performance (perception accuracy). Our method outperforms the separate coding methods, especially in the low signal-to-noise ratio (SNR) regimes. Our main contribution can be summarized as:
\begin{itemize}
	\item To the best of our knowledge, this is the first time a joint source-channel coding architecture is trained for cooperative perception with intermediate fusion over wireless channels. Our method exploits the semantic information through the importance map to achieve better semantic performance and reduce communication overhead. The extracted semantic information is then passed to an autoencoder based on the convolutional neural network (CNN) to overcome the distortion caused by the wireless channel.
        \item  We extend our proposed JSCC architecture to the challenging time-varying multipath fading channels. OFDM blocks combined with channel estimation and channel equalization are adopted to combat the time-varying multipath fading and enhance the system performance. 
	\item Through evaluating the proposed method over AWGN channel, Rayleigh fading channel, and a time-varying multipath fading channel based on the 3GPP specifications \cite{3gpp}, our proposed JSCC communication scheme outperforms the traditional separate source and channel coding methods. Moreover, our proposed method avoids the 'cliff effect' in the low SNR regimes, which prevents catastrophic perception performance reduction. 
        \item Through a robustness study, we verify that only a few parts of the feature map are essential for the intermediate fusion and 3D detection, which demonstrates that the communication bandwidth can be reduced significantly. Moreover, our proposed method can learn robust coded representations of the semantic information that are resilient to various channel models even if our proposed model is trained under one specific channel model.
\end{itemize}

The rest of this paper is organized as follows. Section 2 introduces the system model and problem formulation, including different channel models. Section 3 describes our proposed method, module designs, and training strategies. Section 4 demonstrates the superiority of the proposed model and conducts a robustness study. Section 5 concludes this paper.

\section{System Model and Problem Formulation}
We consider that one CAV would share data with an ego car as an illustrative example, where the shared data is fused with ego cars for 3D detection. The intermediate fusion is considered in our system to achieve a balance between the performance and communication resources. 
\begin{figure*}[htp]
    \centering
    \includegraphics[width=16cm]{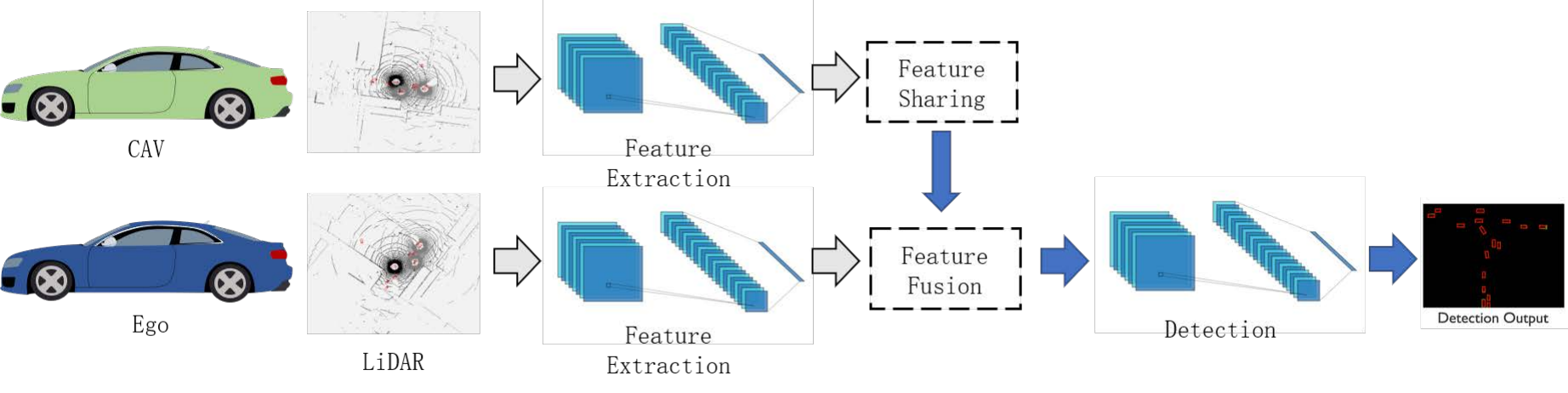}
    \caption{Framework of the cooperative perception with intermediate fusion.}
    \label{fig:framework1}
\end{figure*}
Fig. \ref{fig:framework1} illustrates the framework of the cooperative perception process with intermediate fusion. The cooperative perception system with intermediate fusion comprises four main modules: feature extraction, feature sharing, feature fusion, and detection result generation. 

\textbf{Feature extraction.} We have chosen the anchor-based PointPillar method \cite{lang2019pointpillars}, \cite{chen2019cooper}, \cite{zhou2018voxelnet} as the backbone for 3D detection to extract features from LiDAR point clouds. The raw data would be scattered to form a 2D pseudo-image and subsequently passed to the backbone for further processing.

\textbf{Feature sharing.} In this module, the ego vehicle receives feature maps from the CAV after the feature extraction process. The received intermediate features are then passed to the remaining networks within the ego vehicle. However, in real-world scenarios, due to channel impairment, the transmission of feature maps would often suffer from distortion, resulting in performance degradation.

\textbf{Feature fusion.} The received intermediate feature and the feature extracted in the ego car would be fused to a new feature tensor through the attention network for further processing.

\textbf{Detection result generation.} Upon receiving the final fused feature maps, the prediction header would be employed for the tasks of box regression and classification.

We focus on feature sharing and feature fusion, which are impacted by impairments caused by wireless channels to the most extent. We consider two different channel models to describe the channel impairment. First, we consider Rayleigh fading channels as well as simple additive white Gaussian noise (AWGN) channels. If $x$ is sent, the signal received at the receiver will be represented as
\begin{equation}
    y_i=hx_{i}+n,
\end{equation}
where $h$ represents the channel distortion caused by the Rayleigh fading and $n$ represents the noise. Furthermore, we consider an orthogonal frequency-division multiplexing (OFDM) system for a time-varying multipath fading channel. The received signal $y_{i,k}$ for the $i$th OFDM symbol over the $k$th subcarrier $x_{i,k}$ is represented as
\begin{equation}
    y_{i,k}=h_{i,k}x_{i,k}+n,
\end{equation}
where $h_{i,k}$ represents the channel frequency response for the $i$th OFDM symbol over the $k$th subcarrier and and $n$ represents the noise. $h_{i,k}$ can be represented as
\begin{equation}
    h_{i,k}=\sum_{m=0}^{M-1} a_{m}(i)e^{-j2\pi k\Delta f \tau_m},
\end{equation}
where $M$ represents the number of taps, $\Delta f$ represents the subcarrier spacing, and $(a_{m}(i),\tau_m)$ represent the amplitude and delay of the $m$th channel tap.

For the given channel models, our goal is to develop an encoder and decoder to share the intermediate feature over the wireless channel and optimize the cooperative perception performance. We consider the conventional communication method as the benchmark, which overcomes the distortion caused by the wireless channel through source coding, channel coding, and modulation. However, the traditional method may suffer from the 'cliff effect' in the sense that the perception accuracy drops dramatically when the SNR falls below a threshold. In contrast, we propose to use the autoencoder-based end-to-end learning structure to transmit the intermediate features at the semantic level in order to achieve a desirable balance between communication load and perception accuracy. 

\section{Algorithm based on Importance Map}
\begin{figure*}[htp]
    \centering
    \includegraphics[width=16cm]{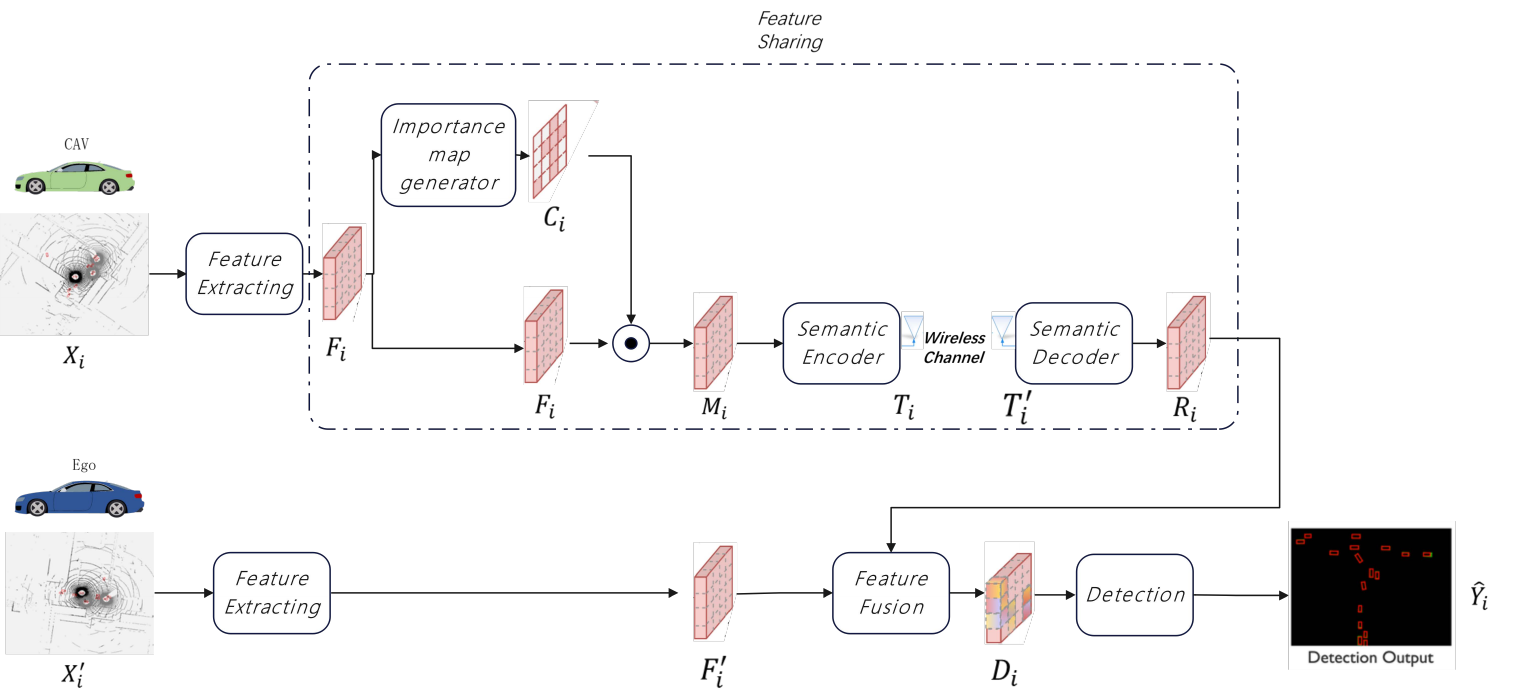}
    \caption{Structure of the cooperative perception model based on the importance map.}
    \label{fig:framework2}
\end{figure*}

In this section, we first introduce the cooperative perception model based on the importance map. Next, we describe the network structure for our proposed semantic encoder and decoder. Then, We extend the cooperative perception model to an OFDM-based system to efficiently deal with the time-varying multipath fading channel. Finally, we illustrate the loss function and training algorithm of our proposed method.

\subsection{Cooperative Perception Model based on Importance Map}
Fig. \ref{fig:framework2} illustrates the structure of the cooperative perception model with the importance map. As demonstrated before, our proposed system focuses on feature sharing and feature fusion, where the data is transmitted over the wireless channel. As illustrated in Fig. \ref{fig:framework2}, feature tensor $F_i$, representing the semantic information of the $i$th sample, is extracted from the raw data $ X_i$  (LiDAR point clouds) in the CAV through a backbone network, such as PointPillar \cite{liu2020when2com},
\begin{equation}
    F_i=\Phi(X_i),
\end{equation}
where $F_i\in{\mathbb{R}^{C\times{H}\times{W}}}$ and $C$ represents the number of channels. $H$ and $W$ are the height and width of the feature tensor. Similarly, the feature tensor of ego car $F^{'}_i$ is also extracted from raw data $X^{'}_i$ in the ego car through the same neural network $ \Phi(\cdot)$, given by
\begin{equation}
    F^{'}_i=\Phi(X^{'}_i).
\end{equation}

For the purpose of more efficient transmission, we endeavor to identify the significant parts of the feature tensor and leverage the importance map \cite{hu2022where2comm}. Specifically, feature tensor $F_i$ is processed through an importance map generator based on neural network $ P(\cdot)$ to obtain an importance map $C_i$, after which $F_i$ will be element-wise multiplied by the generated importance map $C_i$ to generate transmitted data $M_i$, denoted by
\begin{equation}    
    C_i=P(F_i),
\end{equation}
and
\begin{equation}
    M_i={F_i}\odot{C_i}, 
\end{equation}
where $C_i\in{[0,1]^{{H}\times{W}}}$ and $M_i\in{\mathbb{R}^{C\times{H}\times{W}}}$. In this paper, we define the non-zero element ratio of the importance map as compression rate (CR), which indicates the data size of the transmitted information deemed as important. Intuitively, CR is a hyperparameter. Through extensive experiments, we can find that CR could achieve a balance between the transmission overhead and performance when set on the order of $10^{-2}$. Hence, after the element-wise multiplication, the feature tensor becomes sparse and only the non-zero part of the tensor needs to be communicated. Then, we develop a joint source-channel coding scheme for transmitting and receiving the shared data. Unlike traditional communication systems that require source coding, channel coding, and modulation, our method integrates these processes to optimize the transmission of semantic feature tensors. Semantic information $M_i$ would be mapped into a complex symbol stream $T_i$ in the semantic encoder to overcome channel distortion and noise. Complex symbol stream $T_i$ can be represented as
\begin{equation}
    T_i=\Psi_s(M_i),
\end{equation}
where $T_i\in{\mathbb{C}^{H^{'}\times{W^{'}}}}$ and $\Psi_s(\cdot)$ represents the semantic encoder. Following the encoding operation, joint source-channel coded sequence $T_i$ is sent over the communication channel defined via (1) or (2).

At the receiver side (the ego car), received symbol $T^{'}_i$ is first passed to semantic decoder $\Psi_d(\cdot)$, which demaps the complex symbol into semantic information $R_i$ for further fusion, denoted by
\begin{equation}
    R_i=\Psi_d(T^{'}_i).
\end{equation}
Through feature fusion methods, such as attention mechanisms, the semantic information $R_i$ is fused with $F^{'}_i$ into a single fusion tensor that contains the semantic information from the CAV and the ego car. The output of the feature fusion $D_i$ can be represented as
\begin{equation}
    D_i=\chi({R_i,F^{'}_i}), 
\end{equation}
where $D_i\in{\mathbb{R}^{C\times{H}\times{W}}}$, $R_i\in{\mathbb{R}^{C\times{H}\times{W}}}$, and $\chi(\cdot)$ represents the feature fusion network. We remark that the self-attention fusion method is adopted in this paper. Since feature vectors in the feature map, $R_i$ and $F'_i$, correspond to specific spatial regions in the LiDAR point clouds, the spatial relationship can be captured through the self-attention fusion.

After the attention fusion, the prediction header will generate bounding box proposals and their corresponding confidence scores. Detection output $\hat{Y}_i$ generated by the detection network $\Gamma(\cdot)$ can be represented as
\begin{equation}
    \hat{Y}_i=\Gamma(D_i),
\end{equation}
where $\hat{Y}_i$ consists of the regression output and classification output. The regression output consists of seven parameters $(x, y, z, w, l, h, \theta)$, representing the position, the size, and the yaw angle of the predefined anchor boxes, respectively. These parameters are used to refine the anchor boxes and fit the objects accurately. On the other hand, the classification output is the confidence score assigned to each anchor box, indicating the probability of an object or background. The score helps distinguish whether an anchor box contains an object of interest or is just background. Together, the regression and classification outputs compose the final predictions made by the object detection model for object detection and localization.

\subsection{Semantic Encoder and Decoder}
\begin{figure*}[htp]
    \centering
    \includegraphics[width=16cm]{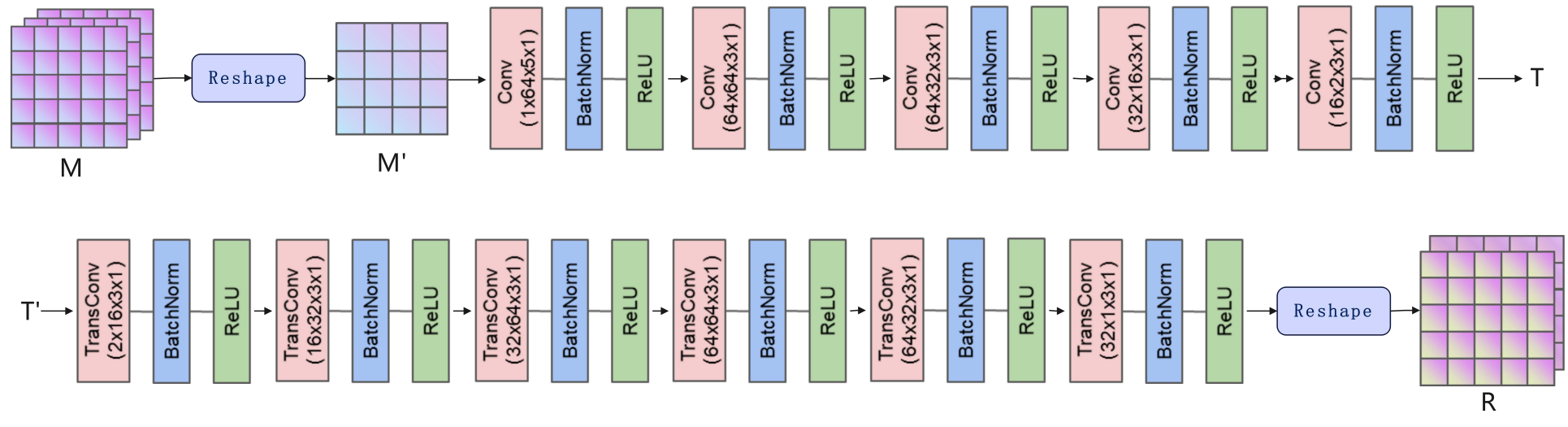}
    \caption{Network structure for our proposed semantic encoder and decoder. The parameters for blue boxes are in the format of $ input ~channel ~size \times output~ channel ~size \times kernel~ size \times stride$.}
    \label{fig:network}
\end{figure*}

The network structure for our proposed semantic encoder and decoder is shown in Fig. \ref{fig:network}. At the transmitter, the semantic encoder transforms input feature map $M\in{\mathbb{R}^{C\times{H}\times{W}}}$ into complex-valued channel input samples $T\in{\mathbb{C}^{H'\times{W'}}}$. This transformation is achieved by using a deterministic encoding function $\Psi_s: {\mathbb{R}^{C\times{H}\times{W}}} \rightarrow {\mathbb{C}^{H'\times{W'}}}$, which satisfies the average power constraint. Firstly, feature map $M$ is reshaped to $M'\in{\mathbb{R}^{1\times{H'}\times{W'}}}$, which neglects the zero part of the feature. Since the magnitude of CR is set as $10^{-2}$, $H'W'$ is much smaller than $CHW$, thus giving rise to a dramatic reduce in both the transmission overhead and the computational complexity. Afterward, the $M'$ is mapped to $T$ with a CNN, which consists of a series of convolutional layers, normalization layers, and parametric ReLU activation functions.

At the receiver, the semantic decoder demaps received complex-valued channel output samples $T'\in{\mathbb{C}^{H'\times{W'}}}$ into reconstructed feature map $R\in{\mathbb{R}^{C\times{H}\times{W}}}$. This demapping process is performed using a deterministic decoding function $\Psi_d: {\mathbb{C}^{H'\times{W'}}} \rightarrow {\mathbb{R}^{C\times{H}\times{W}}}$. In this paper, we apply the transconvolutional layers to reconstruct the received signal, which is more complex than the convolutional layers at the encoder. After the transconvolutional layers, a reshape layer is applied to map the reconstructed signal into $R$. Note that the number of layers in the encoder and decoder can be optimized through experiments.

\subsection{Cooperative Perception Model with OFDM extension }
We extend the cooperative perception model to an OFDM-based system to deal with the time-varying multipath fading channel. In our assumption, each frame of LiDAR point clouds is transmitted within a single time slot, which includes $N_p$ pilot OFDM symbols and $N_s$ information OFDM symbols. We use block-type pilots for channel estimation by sending known symbols on all subcarriers over a few OFDM symbols. Complex symbol stream $T_i$ is initially normalized to $T_n$ according to
\begin{equation}
T_n = \sqrt{H^{'}W^{'}P}\frac{T_i}{\sqrt{T_i^{*}T_i}},
\end{equation}
where $T_n \in{\mathbb{C}^{H^{'}\times{W^{'}}}}$, $T_i^{*}$ is the conjugate transpose of $T_i$, and $P$ is the average transmit power constraint. $T_n$ would be reshaped to $T_m \in{\mathbb{C}^{N_s\times{L_{\text{fft}}}}}$, where $L_{\text{fft}}$ denotes the number of subcarriers in an OFDM symbol. 

In case $H^{'}W^{'}<N_sL_{\text{fft}}$, we would reshape $T_n$ into $T_m$ through zero-padding. The pilot symbols $T_p\in{\mathbb{C}^{N_p\times{L_{\text{fft}}}}}$ are used for channel estimation, which are known to both the transmitter and receiver. In this paper, we assume that the $N_p$ pilot symbols are generated by the QAM modulation symbols of randomly generated bits. After that, we apply the inverse fast Fourier transform (IFFT) to generate $T_p$ and $T_m$ and add the cyclic prefix (CP) to them. Subsequently, the transmit signal $E\in{\mathbb{C}^{(N_p+N_s)\times{(L_{\text{fft}}+L_{\text{cp}})}}}$ travels through the multipath channel as described in  (2). Upon receiving the noisy channel output $E'$, the receiver removes the CP and performs the fast Fourier transform (FFT) to generate the frequency domain pilots $T'_p$ and data symbols $T'_m$. Thus, the received pilot and information symbols can be represented as
\begin{equation}
    T'_m[i,k]=H[i,k]T_m[i,k]+Z,
\end{equation}
and
\begin{equation}
 T'_p[j,k]=H[j,k]T_p[j,k]+Z,
\end{equation}
where $H[j,k]$ denotes the channel frequency response at the $k$th subcarrier of the $j$th OFDM symbol, and $Z$ denotes the AWGN. Meanwhile, $T'_m[i,k]\in{\mathbb{C}^{N_s\times{L_{\text{fft}}}}}$ and $T_p\in{\mathbb{C}^{N_p\times{L_{\text{fft}}}}}$ represent the received frequency domain symbols of the pilots and information, respectively.

One feasible method for estimating the source information is to aggregate $T'_p$, $T'_m$, and $T_p$ and directly input them into a neural network as proposed in \cite{haoye}. However, this approach relies on the neural network and treats the entire decoder as a black box, which may converge slowly. Hence, we apply explicit signal processing kernels, such as channel estimation and equalization, to reconstruct the source information. In this work, we adopt the minimum mean-squared error (MMSE) channel estimation method. Estimated channel response ${H}_{\text{MMSE}}[j]$ can be represent as
\begin{equation}
{H}_{\text{MMSE}}[j]=R_{HH_{\text{LS}}}\left(R_{HH}+\frac{1}{\text{SNR}}I\right)^{-1}H_{\text{LS}},
\end{equation}
where $R_{HH_{LS}}$ represents cross-correlation matrix between real channel response of the $j$th pilot symbol $H$ and least squares (LS) channel estimation $H_{\text{LS}}$, and $R_{HH}$ represents the autocorrelation matrix of $H$. Meanwhile, SNR denotes signal-to-noise ratio of the signal and $I$ represents the identity matrix, which is in the same shape as $R_{HH}$. For simulation, we use $H_{\text{LS}}$ to approximate $H$ when computing $R_{HH}$ and $R_{HH_{\text{LS}}}$ because the real channel response cannot be exactly known at the receiver. The LS channel estimation of the $j$th pilot symbol $H_{\text{LS}}$ can be represented as
\begin{equation}
    H_{\text{LS}}=\frac{T'_p[j]}{T_p[j]},
\end{equation}
where $T_p[j]$ represents the $j$th transmitted pilot symbol and $T'_p[j]$ represents the $j$th received pilot symbol. For equalization, we adopt a conventional MMSE equalizer, which can be represented as
\begin{equation}
    T'[i,k]=\frac{H_{\text{MMSE}}[i,k]^{*}T'_m[i,k]}{|H_{\text{MMSE}}[i,k]|^2+\sigma^2},
\end{equation}
where $H_{\text{MMSE}}[i,k]$ represent the estimated channel response of the $i$th OFDM symbol over the $k$th subcarrier, and $\sigma^2$ represents the noise power.

\subsection{Loss function and Training Algorithm}

By optimizing the reconstruction loss, the semantic encoder and decoder can minimize the average distortion between feature map $M$ and its reconstruction $R$ produced by the decoder. The reconstruction loss can be represented by the average mean-squared error (MSE) between $M$ and $R$,
\begin{equation}
    L_{rec}=\frac{1}{N}\sum_{i=1}^N ||M_i-R_i||_2^2,
\end{equation}
where $i$ represents the $i$th sample and $N$ denotes the number of samples. Meanwhile, in order to measure the performance of cooperative perception, we adopt smooth $L_1$ loss for regression and a focal loss for classification \cite{xu2022opv2v}, denoted by
\begin{equation}
    L_{per}=\frac{1}{N}\sum_{i=1}^N L(Y_i,\hat{Y}_i),
\end{equation}
where $L(\cdot)$ denotes the perception loss for one sample. The total loss function can be represented by
\begin{equation}
    L_{total}=\lambda_1 L_{rec} + \lambda_2 L_{per},
\end{equation}
where $\lambda_1$ denotes the weight for the reconstruction loss and $\lambda_2$ denotes the weight for the perception loss. 
Minimizing $L_{rec}$ amounts to encouraging the transmitted and recovered messages to be as similar to each other as possible, and can thus be roughly characterized as improving the bit-level accuracy. On the other hand, minimizing $L_{per}$ drives down the perception loss and therefore can be seen as the semantic-level optimization. Hence, the system can achieve a balance between semantic and bit accuracy through choosing appropriate hyperparameters $\lambda_1$ and $\lambda_2$.

\begin{algorithm2e}
\SetAlgoLined
\KwIn{The raw data $ X_i$  (LiDAR point clouds)}
\KwOut{The detection output $\hat{Y}_i$}
 \textbf{Step 1}: Train the semantic encoder and decoder. 
 \\
 \While{ not converge}{
   \For{$X_i$ in batch samples}{
         $F_i=\Phi(X_i)$\;
         $M_i={F_i}\odot{Ci}, C_i=P(F_i)$\;
         $T_i=\Psi_s(M_i)$\;
         $R_i=\Psi_d(T^{'}_i)$\;
   }
   Compute $L_{rec}$\;
   Compute the gradient\;
   Update the network $\Psi_s(\cdot)$,$\Psi_d(\cdot)$ using SGD;
   }
\textbf{Step 2}:  Train the whole network. 
 \\
 \While{ not converge}{
   \For{$X_i$ in batch samples}{
         $F_i=\Phi(X_i)$\;
         $F^{'}_i=\Phi(X^{'}_i)$\;
         $M_i={F_i}\odot{Ci}, C_i=P(F_i)$\;
         $T_i=\Psi_s(M_i)$\;
         $R_i=\Psi_d(T^{'}_i)$\;
         $D_i=\chi({R_i,F^{'}_i})$\;
         $\hat{Y}_i=\Gamma(D_i)$\;
   }
   Compute $L_{total}$\;
   Compute the gradient\;
   Update the whole network using SGD\;
   }

 \caption{Training Algorithm}
\end{algorithm2e}

As illustrated in Algorithm 1, the training process of the proposed model consists of two steps. The first step is to train the semantic encoder and decoder until convergence. Particularly, we first generate $R_i$ and compute the loss between $R_i$ and $F_i$ in   (18). Then, we compute the gradient and update semantic encoder $\Psi_s(\cdot)$ and decoder $\Psi_d(\cdot)$ using stochastic gradient descent (SGD) methods. As for the second step, the detection output $\hat{Y}_i$ is generated and passed through the whole network to compute the loss in  (20). Finally, the whole network would be updated until convergence. It is necessary to adopt the first step when training since training the whole network directly may lead to slow convergence or even divergence. Moreover, the second step would enable the whole network to perform well at the semantic level.

\section{Evaluation}
To demonstrate the potential of our proposed cooperative perception model with the importance map, we evaluate the architecture depicted in Fig. \ref{fig:network}. Our proposed model is evaluated on the OPV2V dataset \cite{xu2022opv2v}, which is a vehicle-to-vehicle cooperative perception dataset co-simulated by OpenCDA \cite{xu2022opv2v} and Carla \cite{dosovitskiy2017carla}. The dataset includes 12K frames of 3D point clouds and RGB images with 230K annotated 3D boxes. The perception range is $40m\times40m$. For Lidar-based 3D object detection task, our proposed model leverages PointPillar backbone \cite{lang2019pointpillars}, which transforms the field of view into a bird’s eye view map. The whole network is implemented in Pytorch and trained with two RTX 3080 GPUs. We employ the Adam optimization framework for back-propagation, which represents a variant of stochastic gradient descent.

The performance of our proposed model is evaluated in terms of Average Precision (AP) at the Intersection-over-Union (IoU) threshold of 0.50 and 0.70. AP is a commonly used evaluation metric to assess the performance of object detection algorithms. It quantifies the accuracy and trade-off between precision and recall by calculating precision at different recall levels and taking their average. A higher AP value indicates that the algorithm is more accurate and reliable in detecting objects. The AP is defined as
\begin{equation}
    \text{AP}@I=\int_{0}^{1} \text{max}\left\{p(r'|r'>r)\right\}dr,
\end{equation}
where $p(r)$ is the precision-recall curve at IoU threshold $I \in \left\{0.5,0.7\right\}$. For instance, AP$@0.5$ represents the area under the precision-recall curve at IoU threshold 0.5. Meanwhile, the performance of AP$@0.5$ is intuitively better than AP$@0.7$ since the higher IoU threshold would decrease the precision accuracy.

The channel SNR can be defined as
\begin{equation}
    \text{SNR}=10\log_{10} \frac{P}{{\sigma}^2},
\end{equation}
where SNR represents the ratio of the average power of the coded signal (channel input signal) to the average noise power. In the proposed scheme, $P$ denotes the average power of the channel input signal after the power normalization layer applied at the encoder, and ${\sigma}^2$ represents the average noise power. For the benchmark schemes that utilize explicit signal modulation, $P$ refers to the average power of the symbols in the constellation. To ensure generality, the average signal power is set to $P = 1$ for all experiments.
\begin{figure*}[htbp]
    \begin{minipage}[t]{0.5\linewidth}
        \centering
        \includegraphics[width=\textwidth]{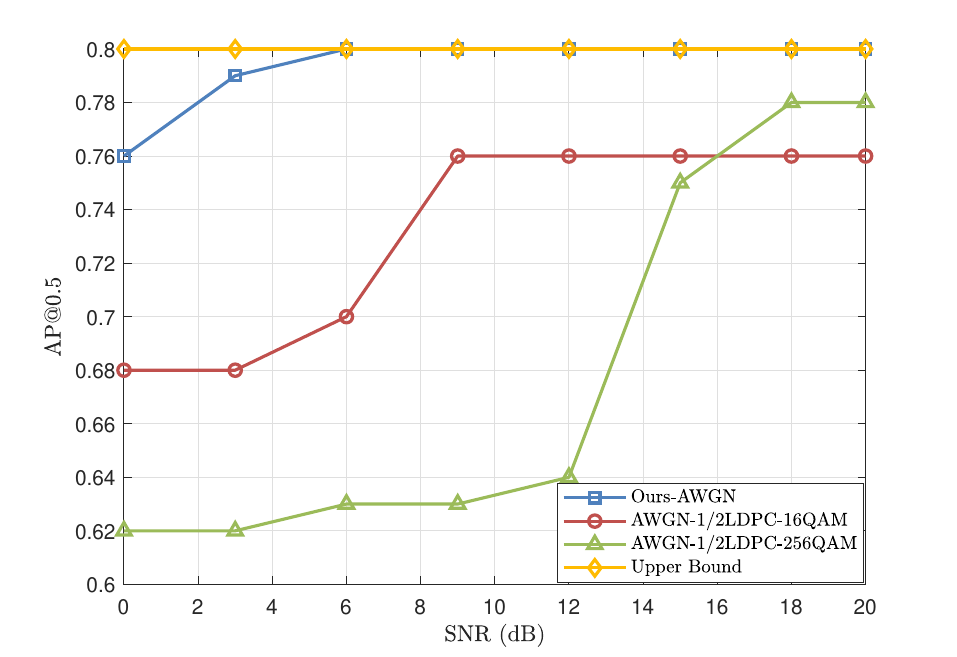}
        \centerline{(a)}
    \end{minipage}%
    \begin{minipage}[t]{0.5\linewidth}
        \centering
        \includegraphics[width=\textwidth]{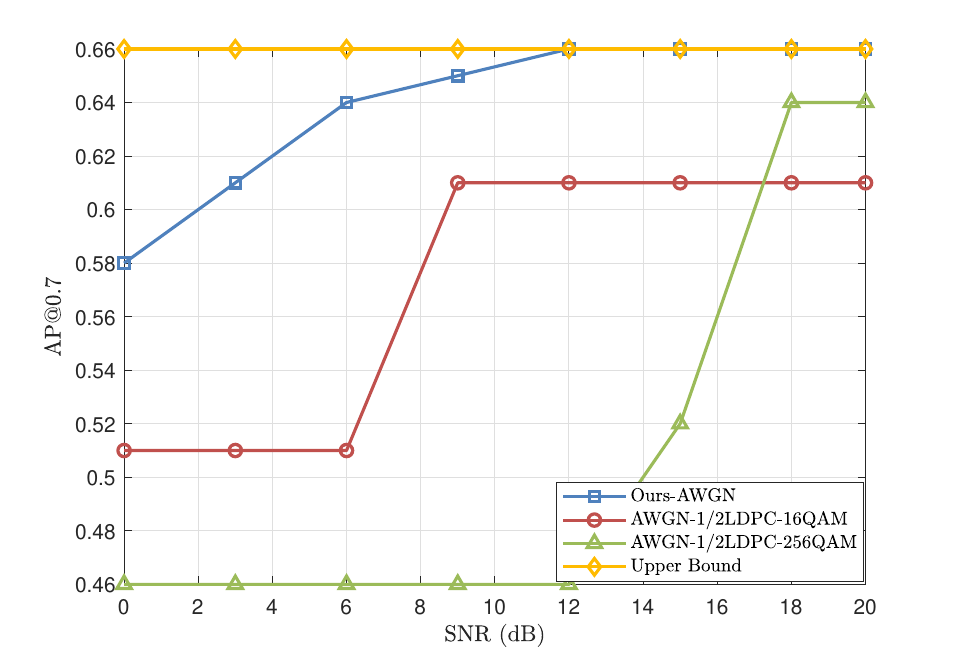}
        \centerline{(b)}
    \end{minipage}
    \caption{Performance of our proposed method compared with baseline schemes at different SNRs over the AWGN channel. The baseline schemes and our proposed method exploit the same channel uses for fairness. (a) Average precision performance at the IoU threshold 0.5; (b) Average precision performance  at the IoU threshold 0.7.}
    \label{fig:AWGN}
\end{figure*}

\subsection{Comparison with  Separate Coding Schemes}
In this section, we compare our proposed method with a baseline separate source and channel coding scheme. The baseline separate coding scheme uses uniform quantization for the feature generated by the backbone network, i.e., $M_i$ in (7). In the baseline scheme, we set the 8-bit quantization method as source coding. Note that the performance of the perception model will not decrease under the 8-bit quantization method by selecting the appropriate quantization step size and zero point. Furthermore, the $1/2$ rate low-density parity check (LDPC) coding with a 1,000 code length are considered for channel coding, and 16QAM or 256QAM is applied for modulation after channel coding.

Meanwhile, different channel coding schemes combined with different modulation methods may cause different channel uses for the same input. For a fair comparison, we adjust the CR of the importance map to ensure the same channel uses for different combinations of modulation and coding schemes (MCS). The size of raw data can be represented by $D_r$, and the channel uses  of baseline schemes can be represented as

\begin{equation}
    \text{Channel uses} = \frac{D_r \times \text{CR}\times8}{log_2{M_c}\times R_c },
\end{equation}
where $R_c$ represents the coding rate and $M_c$ represents the order of QAM. Thus,  $\text{CR}=0.00125$ and $\text{CR}=0.0025$ can be set for 1/2LDPC+16QAM and 1/2LDPC+256QAM, respectively, which can ensure the same channel uses as our proposed methods with $\text{CR}=0.005$.

\begin{figure*}[h]
    \begin{minipage}[t]{0.5\linewidth}
        \centering
        \includegraphics[width=\textwidth]{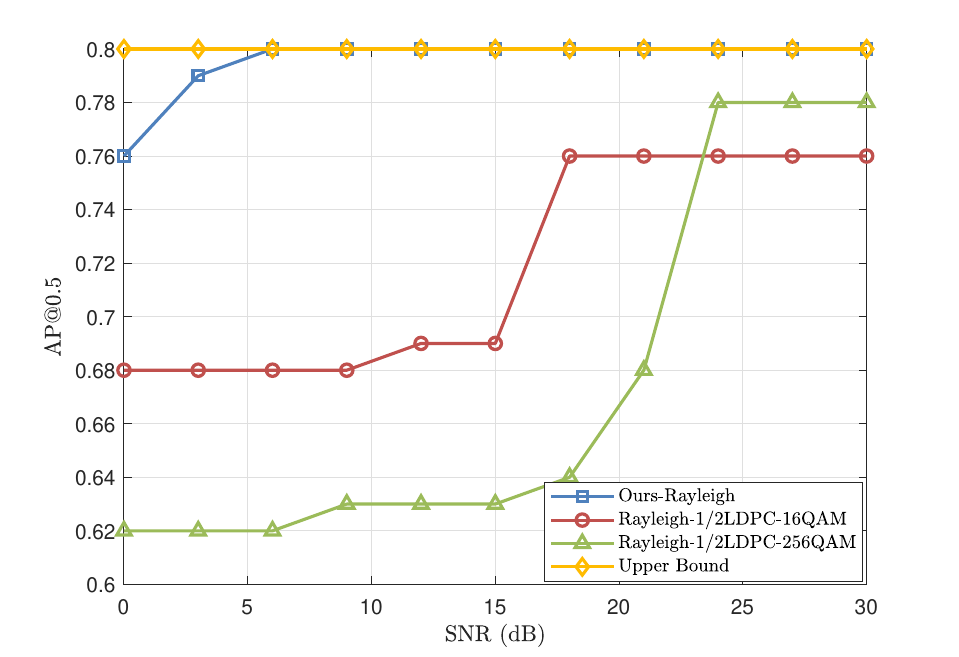}
        \centerline{(a)}
    \end{minipage}%
    \begin{minipage}[t]{0.5\linewidth}
        \centering
        \includegraphics[width=\textwidth]{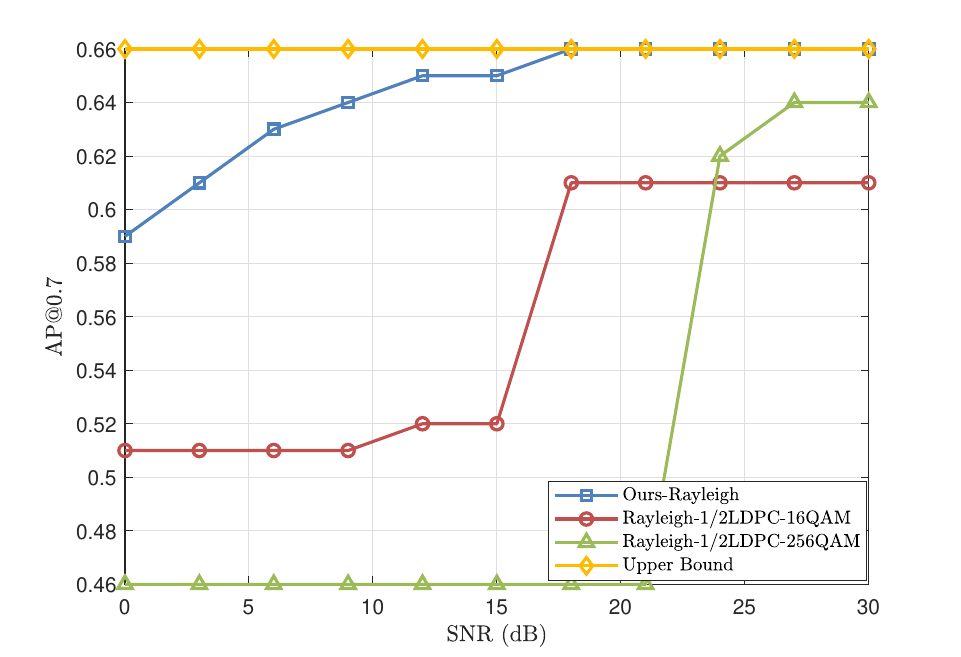}
        \centerline{(b)}
    \end{minipage}
    \caption{Performance of our proposed method compared with baseline schemes at different SNRs over the Rayleigh fading channel. The baseline schemes and our proposed method exploit the same channel uses for fairness. (a) Average precision performance  at the IoU threshold 0.5; (b) Average precision performance  at the IoU threshold 0.7.}
    \label{fig:Rayleigh}
\end{figure*}

Fig. \ref{fig:AWGN} illustrates the performance of our proposed method compared with baseline schemes at different SNRs over the AWGN channel, where Ours-AWGN denotes our scheme. The upper bound can be obtained assuming perfect communication. In Fig. \ref{fig:AWGN}(a), in terms of AP$@0.5$, our proposed method outperforms the separate coding schemes with the same channel uses since our proposed method is designed and trained with joint source-channel coding, which can jointly optimize the whole system to preserve more semantic information given the same communication resources used. Meanwhile, the proposed method exhibits smoother performance in low SNR regimes without the cliff effect, which is commonly observed in traditional communication systems. Hence, our proposed method can prevent catastrophic perception performance loss in the low SNR regimes, which is critical to autonomous driving. Moreover, our method significantly outperforms traditional schemes and approaches the performance upper bound constrained by the cooperative perception module at SNR = 6 dB. On the one hand, traditional coding combined with low-order modulation achieves better performance in the low SNR regimes and reaches its maximum performance at SNR = 9 dB. On the other hand, despite performing worse than low-order modulation in the low SNR regimes, high-order modulation reaches its maximum performance at SNR = 18 dB and surpasses the performance of low-order modulation. This is because high-order modulation allows the transmission of more semantic information, especially in high SNR regimes. In Fig. \ref{fig:AWGN}(b), our proposed method is compared to the traditional method evaluated in terms of AP$@0.7$. We observe that our proposed method still outperforms traditional methods while the performance of AP$@0.7$ drops compared to the AP$@0.5$. The decrease in the performance of AP$@0.7$ compared to that of AP$@0.5$ is because AP at higher IoU would judge the detection outputs more strictly. Hence, AP$@0.7$ is more sensitive to noise than AP$@0.5$. Accordingly, in our proposed methods, the performance of the AP$@0.5$ approaches the upper bound at SNR = 6 dB while the performance of AP$@0.7$ approaches the upper bound at SNR = 12 dB.

Next, the performance of our proposed scheme is considered under the assumption of a slow Rayleigh fading channel with AWGN. In this case, the channel transfer function is $y(x) = hx +n $, where $h\sim{\mathcal{CN}(0,H_c)}$ represents the channel distortion caused by the Rayleigh fading and $n\sim{\mathcal{CN}(0,\sigma^{2})}$ represents the noise. We set $H_c$ as 1 and simulate different channel SNRs through varying $\sigma^2$. In this experiment, we assume perfect channel state information at the receiver, which can be represented by $\hat{y}=\frac{y}{||h||}$. Fig. \ref{fig:Rayleigh}  illustrates the performance of our proposed method compared to baseline schemes at different SNRs over the Rayleigh fading channel, where Ours-Rayleigh denotes the proposed method. From Fig. \ref{fig:Rayleigh}(a) and \ref{fig:Rayleigh}(b), the performance of both ours and baseline schemes decreases compared to the static AWGN channel while the proposed method still outperforms other traditional schemes. Comparing Fig. \ref{fig:AWGN}(b) with Fig. \ref{fig:Rayleigh}(b), our proposed method approaches the upper bound at SNR = 18 dB over the Rayleigh fading channel while reaching the upper bound at SNR = 12 dB over the AWGN channel, which has a gap of 6 dB. This gap remains consistent in other baseline schemes or evaluated in terms of AP$@0.5$. Fig. \ref{fig:Rayleigh} demonstrates that due to the distortion caused by the Rayleigh fading, both the proposed method and baseline schemes approach the upper bounds in the higher SNR regimes than the AWGN channel.

\subsection{Ablation study}

\begin{figure}[htp]
    \centering
    \includegraphics[width=0.5\textwidth]{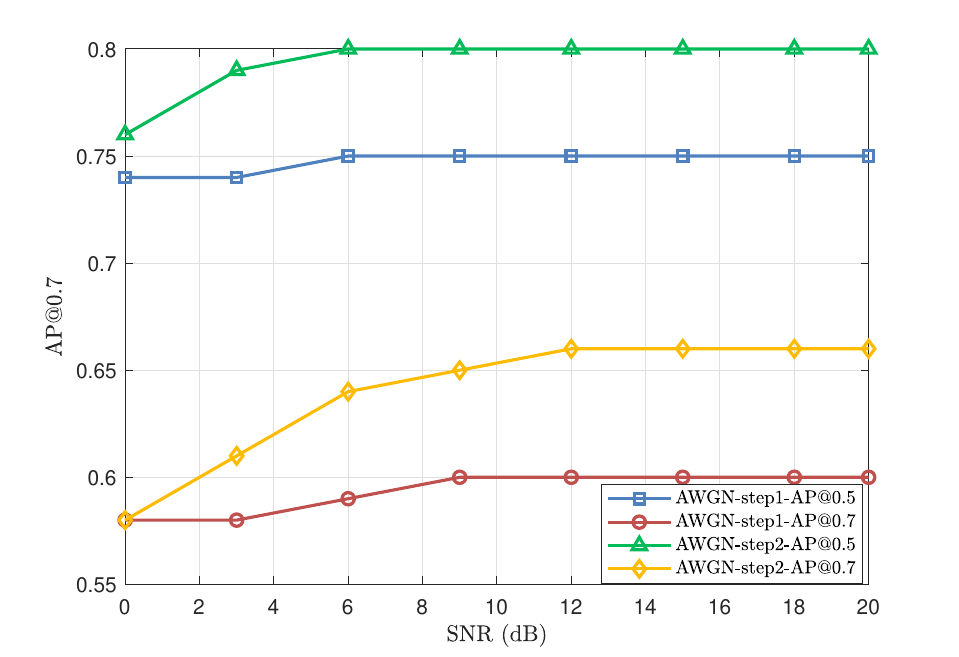}
    \caption{Performance of our proposed method with different training steps over an AWGN channel.}
    \label{fig:step}
\end{figure}

In this section, we investigate the impact of different training steps with respect to the channel SNR over an AWGN channel. As Algorithm 1 illustrates, our training algorithm includes two steps, to train the channel encoder and decoder with the MSE loss in (18) and to train the whole model jointly with loss in (20). From Fig. \ref{fig:step}, the performance of going through 2 training steps evaluated in terms of both AP$@0.5$ and AP$@0.7$ has increased significantly than that training for only Step 1 in Algorithm 1. This is because training with one step can only allow the system to transmit accurately at the bit level while two-step training can reach accuracy at the semantic level. Meanwhile, a joint end-to-end training scheme for task loss (perception loss) can provide valuable insights to the semantic encoder regarding the essential semantic information that needs to be transmitted. Moreover, despite that the MSE loss of different training steps may be roughly the same (e.g, the MSE loss is 0.008 and 0.01 for AWGN-step1-AP@0.5 and AWGN-step2-AP@0.5, respectively), the semantic accuracy between different training steps would have an explicit gap, which demonstrates that the accuracy at the bit level may not correspond to that at the semantic level.
\begin{figure}[t]
    \centering
    \includegraphics[width=0.5\textwidth]{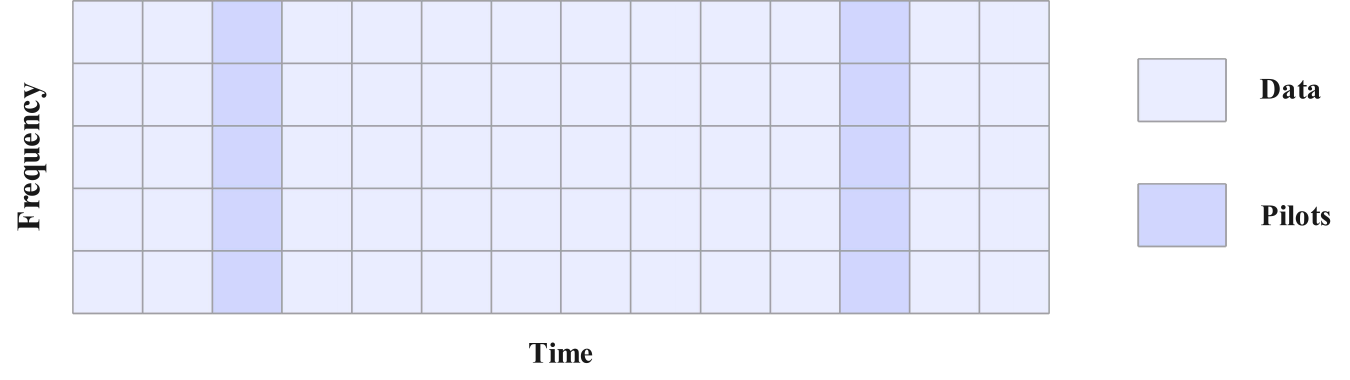}
    \caption{Kronecker-based pilot for channel estimation of a time-varying OFDM channel.}
    \label{fig:pilot}
\end{figure}

\begin{figure*}[h]
    \begin{minipage}[t]{0.5\linewidth}
        \centering
        \includegraphics[width=\textwidth]{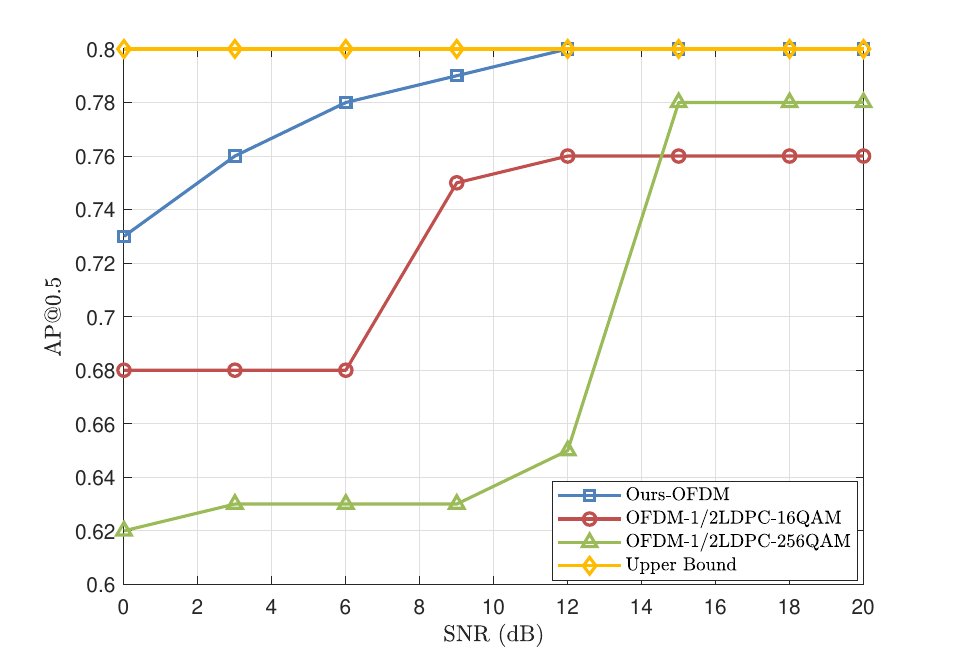}
        \centerline{(a)}
    \end{minipage}%
    \begin{minipage}[t]{0.5\linewidth}
        \centering
        \includegraphics[width=\textwidth]{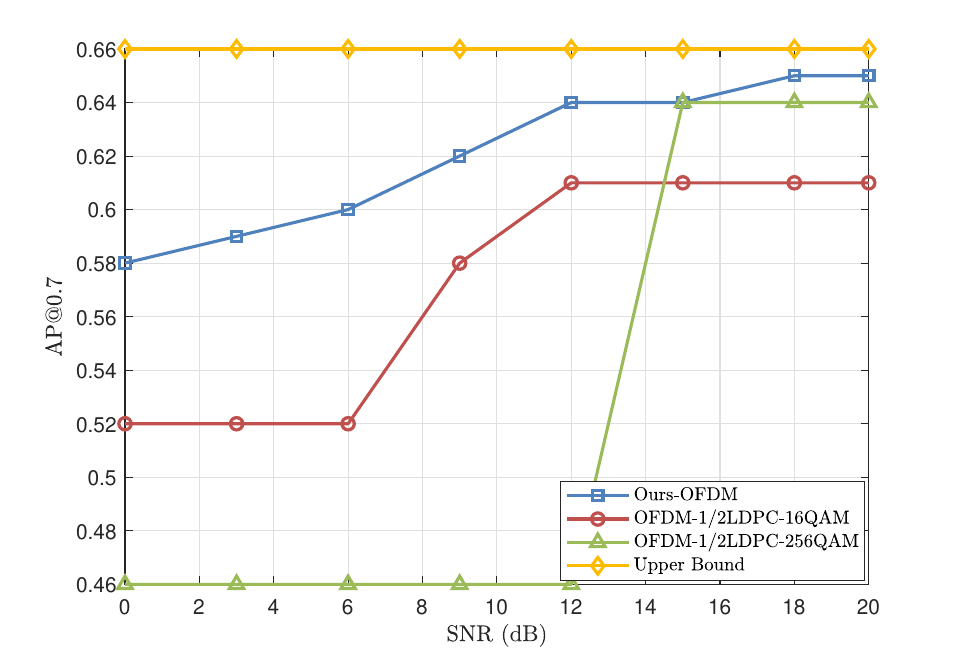}
        \centerline{(b)}
    \end{minipage}
    \caption{Performance of our proposed method compared with baseline schemes at different SNRs over a time-varying multipath fading channel. The baseline schemes and our proposed method exploit the same channel uses for fairness. (a) Average precision performance  at the IoU threshold 0.5; (b) Average precision performance  at the IoU threshold 0.7.}
    \label{fig:OFDM}
\end{figure*}
\subsection{Performance over the Time-varying Multipath Fading Channel}
We train and test our proposed system over a time-varying multipath fading channel, which is modeled based on the specifications provided by 3GPP \cite{3gpp}, where the tapped delay line (TDL) model is used for evaluations. The channel coefficients are generated using a sum-of-sinusoids (SoS) model to simulate the multipath propagation. In addition, we simulate channel aging to account for mobility scenarios, which can be represented by the speed of the users. The power delay profiles (PDPs) of the TDL models are normalized to have a total energy of one for simplified evaluation. We construct three TDL models, namely TDL-A, TDL-B, and TDL-C, to represent different channel profiles for Non-Line-of-Sight (NLOS) conditions. For Line-of-Sight (LOS) scenarios, we construct TDL-D and TDL-E, with their parameters specified in \cite{3gpp}. The Doppler spectrum for each tap in the channel model follows a classical (Jakes) spectrum shape, and the maximum Doppler shift is denoted as $f_D$. In TDL-D and TDL-E, which include a LOS path, the first tap follows a Ricean fading distribution. Additionally, the Doppler spectrum for these taps contains a peak at the Doppler shift $f_S = 0.7 f_D$, with an amplitude adjusted to achieve the specified K-factor for the resulting fading distribution.

The RMS delay spread values of the TDL models are normalized, and they can be scaled in delay to achieve a desired RMS delay spread. The scaled delays can be calculated using the following equation
\begin{equation}
    \tau_{n,\text{model}} = \tau_{n,\text{scaled}}\cdot \text{DS}_{\text{desired}},
\end{equation}
where $\tau_{n,\text{scaled}}$ is the normalized delay value of the $nth$ cluster in a TDL model, $\tau_{n,\text{model}}$ is the new delay value (in ns) of the $n$th cluster, and $\text{DS}_{\text{desired}}$ is the desired RMS delay spread (in ns). The scaling parameters for the example are chosen based on Table \ref{tab:table1}. These values are selected to cover the range of RMS delay spreads typically observed in 5G deployments.

\begin{table}[!ht] 
\centering
\caption{Example scaling parameters for TDL models.}
\begin{tabular}{|c|c|} \hline 
 Model & $\text{DS}_{\text{desired}}$  \\ \hline
 Very short delay spread & 10 ns\\ \hline
Short delay spread & 30 ns\\ \hline
Nominal delay spread & 100 ns \\ \hline
Long delay spread & 300 ns \\ \hline
Very long delay spread & 1000 ns\\ \hline
\end{tabular}
\label{tab:table1}
\end{table}

\begin{table}[!ht] 
\centering
\caption{Simulation parameters for OFDM channel.}
\begin{tabular}{|c|c|} \hline 
 Parameters & Values  \\ \hline
  Number of subcarriers & 2048\\ \hline
Number of OFDM symbols & 14\\ \hline
Subcarrier spacing & 15 kHz \\ \hline
Carrier frequency &  3.5 GHz\\ \hline
$\text{DS}_{\text{desired}}$ &  30/100/300 ns\\ \hline
\end{tabular}
\label{tab:ofdm}
\end{table}

As Table \ref{tab:ofdm} illustrates, we set the number of subcarriers as $L_{\text{fft}}=2048$, the subcarrier spacing as 15 kHz, and the carrier frequency as 3.5 GHz. The number of information symbols $N_s$ is set as 12 and the number of pilots $N_p$ is set as 2, which means that one frame of Lidar point can be transmitted in 14 OFDM symbols. Meanwhile, various values of $\text{DS}_{\text{desired}}$ can be set to simulate different scenarios. Moreover, as for channel estimation, three methods are under consideration, including perfect channel estimation, Kronecker-based pilots estimation, and estimation with one pilot. Fig. \ref{fig:pilot} shows the structure of the Kronecker-based pilots for channel estimation of the proposed time-varying OFDM channel, where the third and the twelfth symbols are reserved for pilot transmission. Hence, the whole channel of 14 OFDM symbols can be estimated through linear interpolation to represent the time-varying channel state information (CSI). Meanwhile, the first OFDM symbol is reserved for pilot transmission when estimating with one pilot.

Fig. \ref{fig:OFDM} illustrates the performance of the proposed method compared with baseline schemes at different SNRs over a time-varying OFDM channel. In this figure, We evaluate both the proposed method and baseline schemes in the TDL-A channel with $\text{DS}_{\text{desired}}=300$ ns. Compared to schemes that use separate channel source coding, our proposed method outperforms the benchmark digital transmission schemes in time-varying channels in very low SNR regimes and very high SNR regimes. While the conventional transmission schemes perform well only in channel conditions for which they have been optimized, our proposed method is more robust to channel quality fluctuations. It is noted that in this time-varying multipath fading channels, all evaluated methods suffer from imperfect channel estimation, in addition to additive noise.  Hence, the performance of our proposed method under the time-varying channel has a  gap with that under the AWGN channel and Rayleigh fading channel, especially in low SNR regimes if we compare the performance with that in Fig. \ref{fig:AWGN} and \ref{fig:Rayleigh}.

\begin{figure}[htp]
    \centering
    \includegraphics[width=0.5\textwidth]{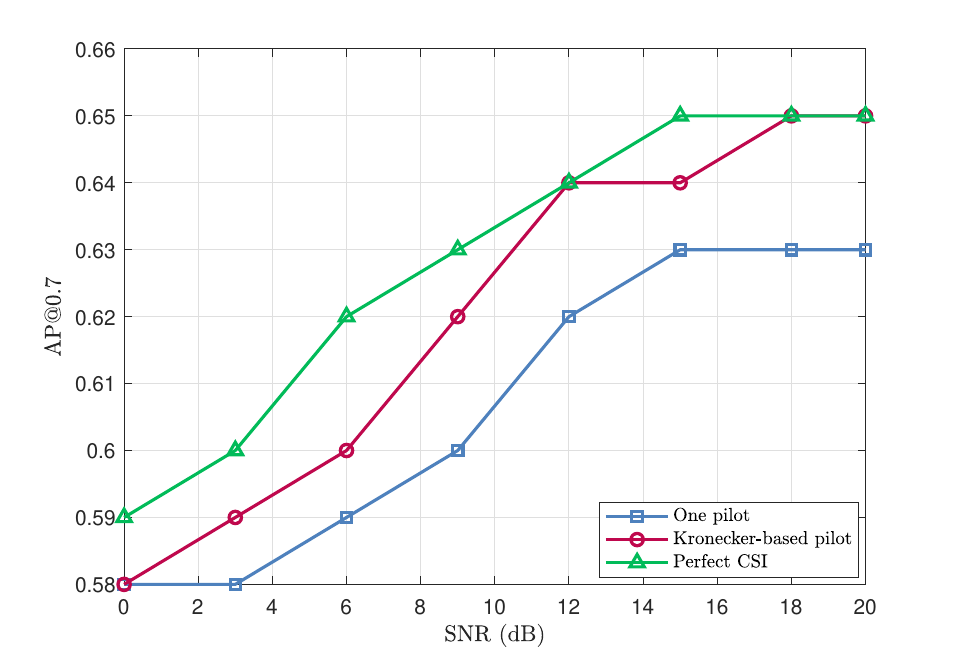}
    \caption{Performance of our proposed method with three various channel estimation schemes.}
    \label{fig:ce}
\end{figure}
\begin{figure}[htp]
    \centering
    \includegraphics[width=0.5\textwidth]{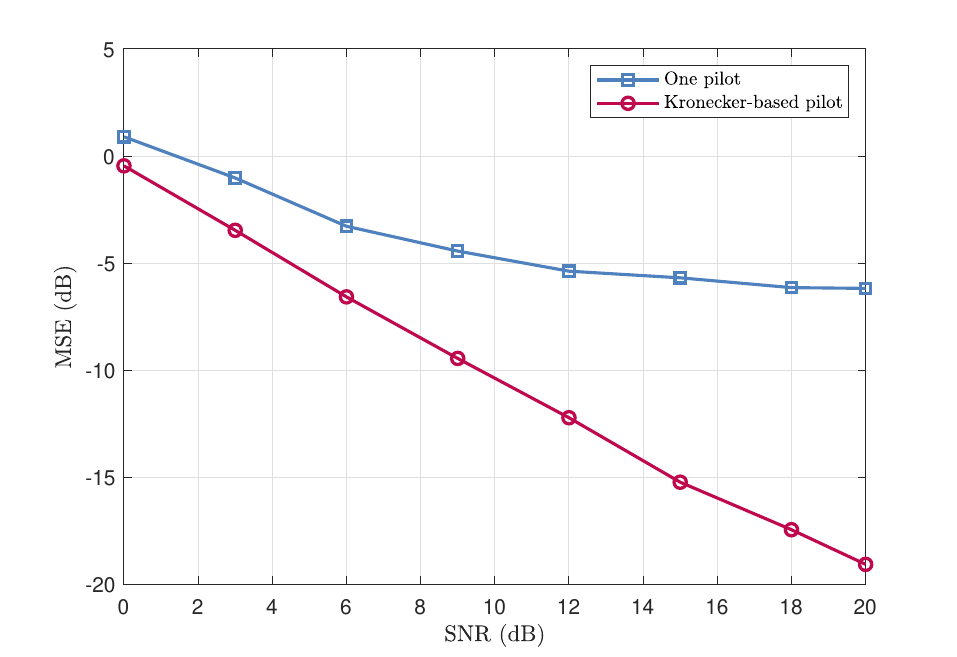}
    \caption{The MSE of various channel estimation schemes.}
    \label{fig:mse}
\end{figure}

\begin{figure*}[t]
    \label{fig:example}
	\centering
	\subfigure[SNR = 0 dB]{
		\includegraphics[width=0.48\textwidth]{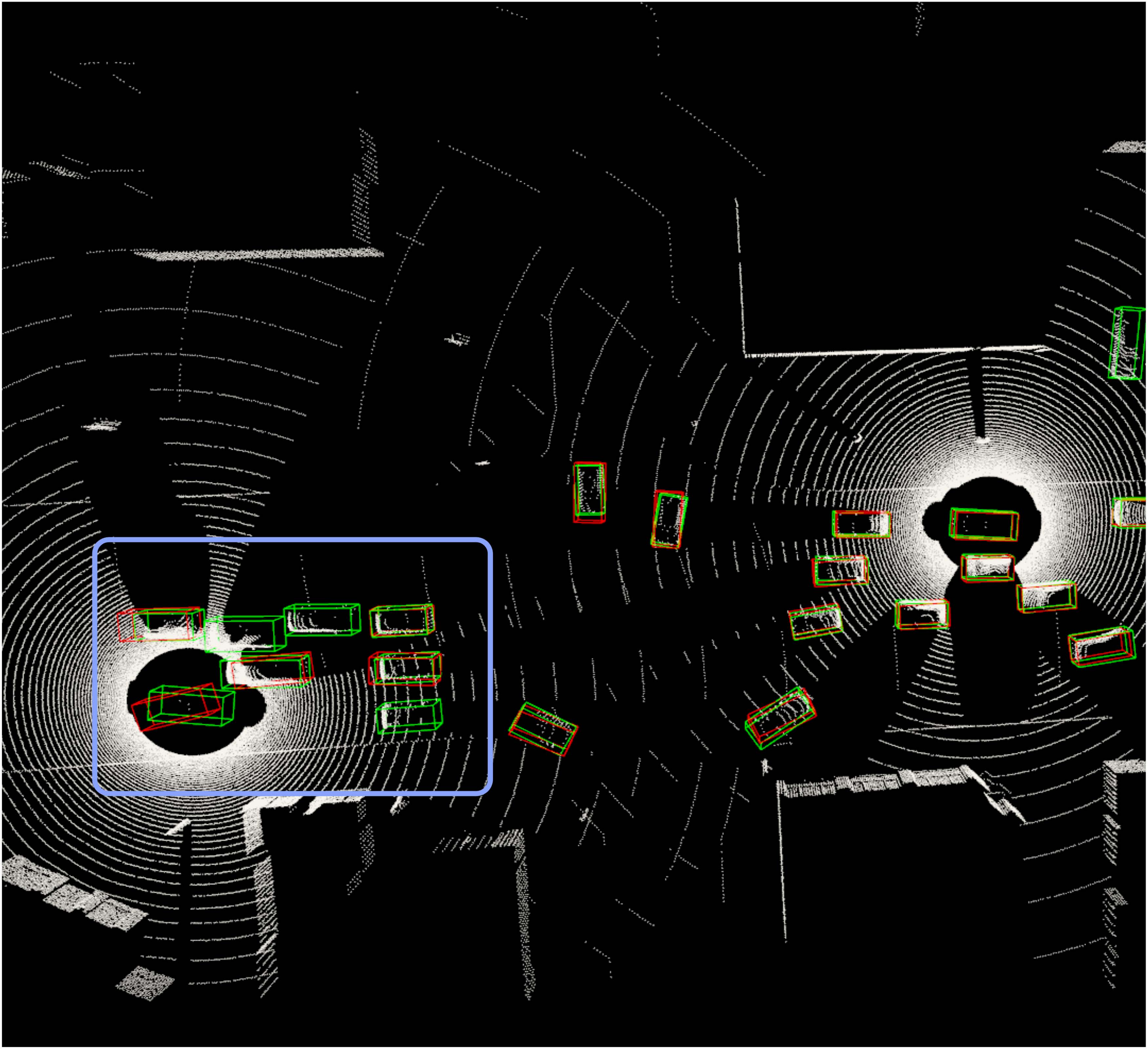}
	}
	\subfigure[SNR = 3 dB]{
		\includegraphics[width=0.48\textwidth]{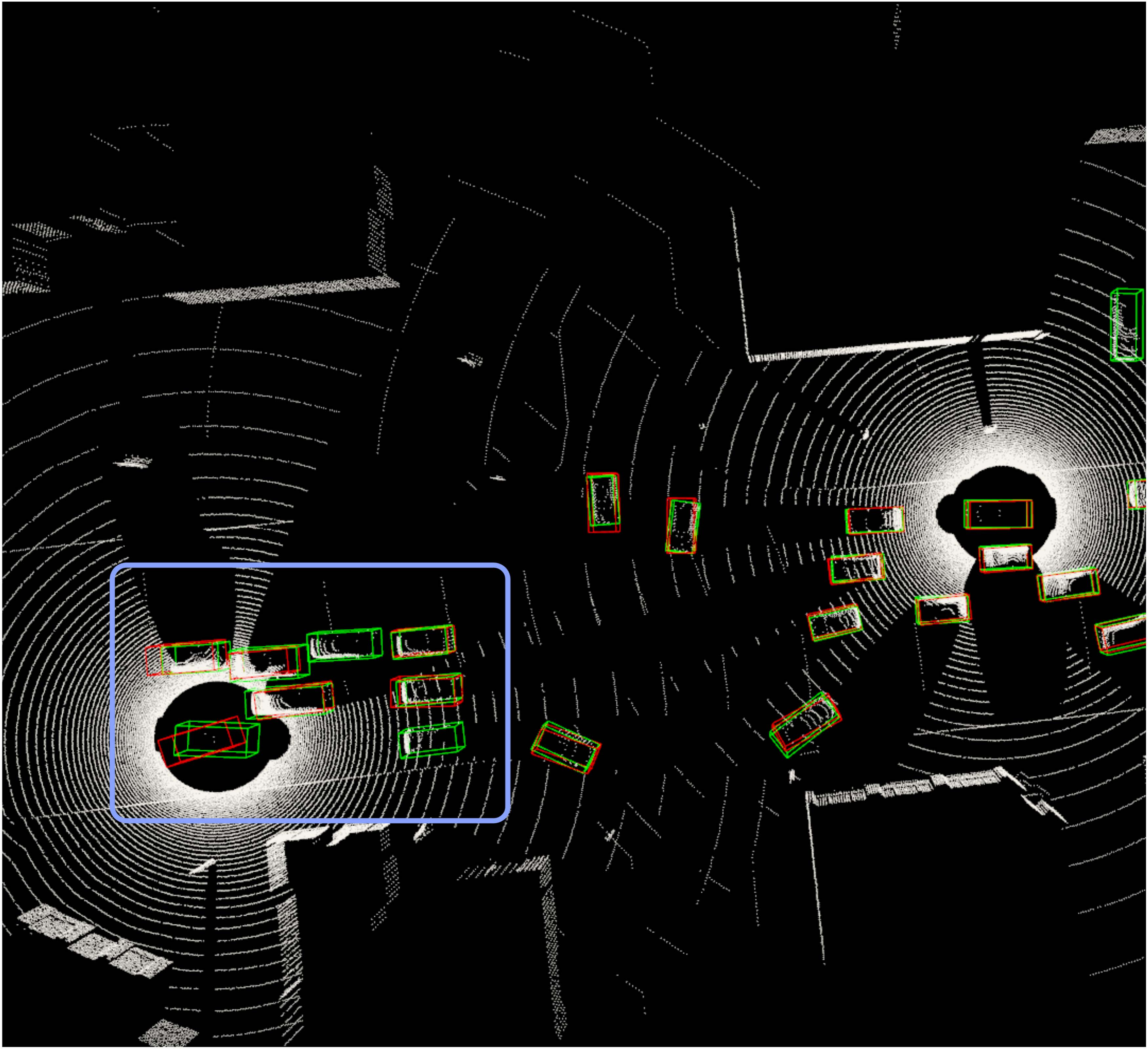} 
	}
        \subfigure[SNR = 10 dB]{
		\includegraphics[width=0.48\textwidth]{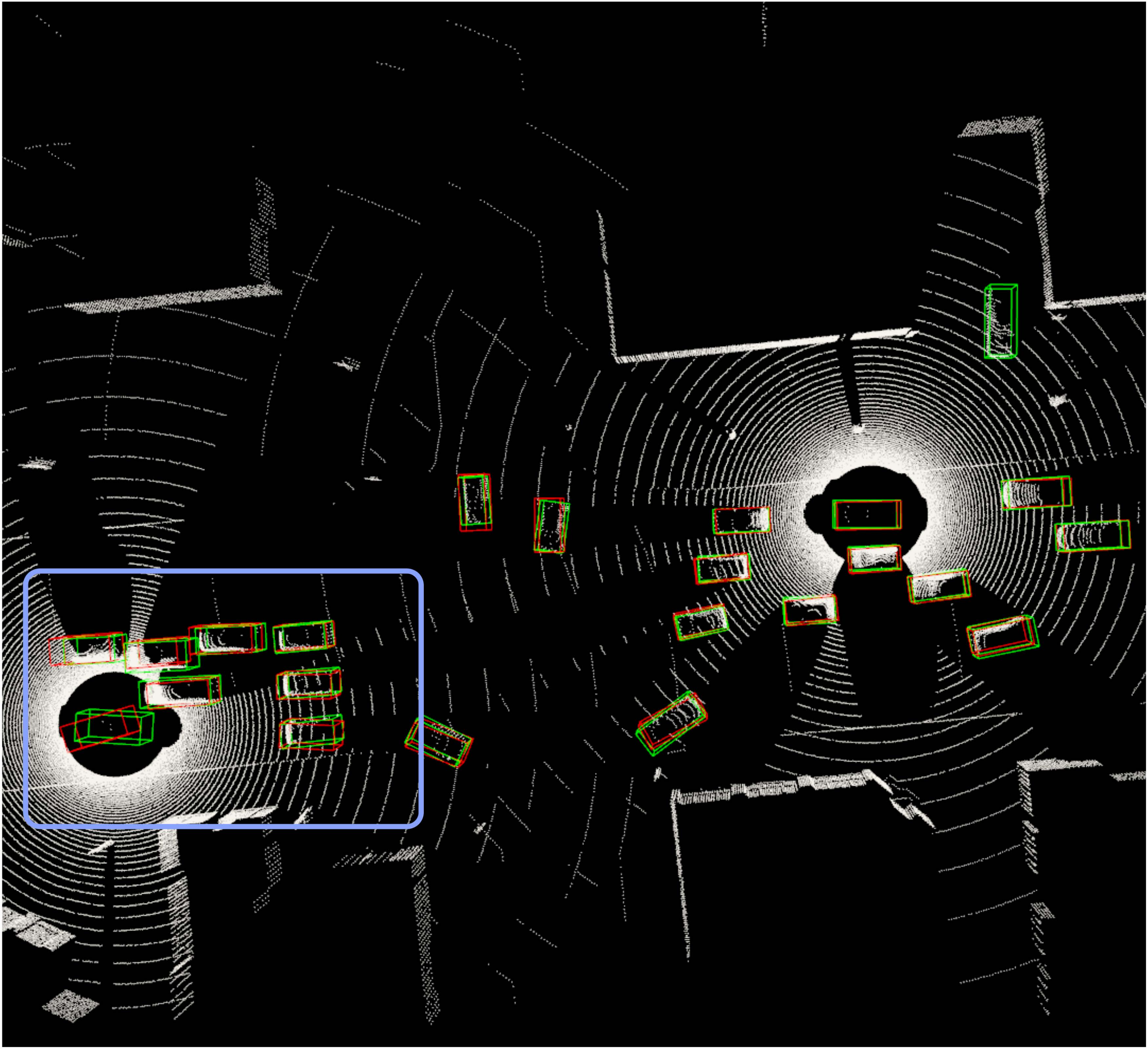}
	}
	\subfigure[SNR = 20 dB]{
		\includegraphics[width=0.48\textwidth]{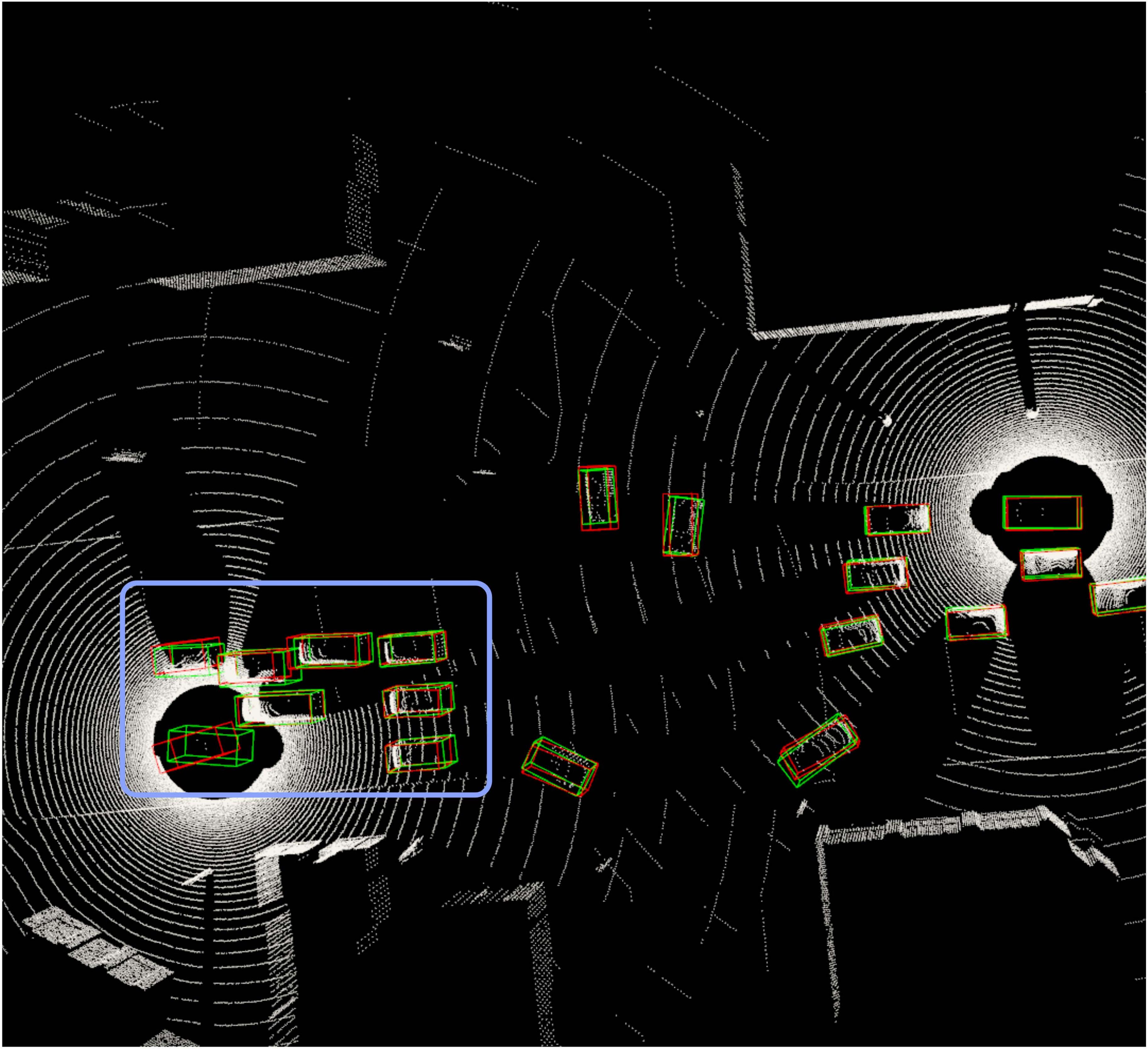} 
	}
	\caption{Examples of 3D detection outputs produced by our proposed method.}
    \label{fig:example}
\end{figure*}
Next, we consider the impact of channel estimation, which depends on specific channel estimation methods. Fig. \ref{fig:ce} illustrates the performance of our proposed method with three different channel estimation schemes while Fig. \ref{fig:mse} shows the corresponding MSE of different channel estimation schemes. The performance evaluated with perfect CSI (labeled as Perfect CSI) serve an upper bound. As Fig. \ref{fig:ce} illustrates, Kronecker-based pilot estimation (labeled as Kronecker-based pilot) outperforms estimation with one pilot (labeled as One pilot). This is because Kronecker-based pilot estimation can estimate the channel more precisely than that with only one pilot over the time-varying channel, especially in the low SNR regimes, which can be shown in Fig. \ref{fig:mse}. Hence, transmitted information would be reconstructed more accurately with more accurate channel estimation results through equalization. This result emphasizes the impact of the channel estimation schemes when communicating over a time-varying multipath fading channel, which further demonstrates the importance of channel estimation design. Finally, a visual example of 3D detection results produced by our proposed method under different SNRs over the multipath fading channels is presented in Figs. \ref{fig:example}. The red boxes represent the predicted detection results, and the green boxes denote the ground truth. Since the transmitted intermediate feature is corrupted severely in low regimes, several vehicles are not detected in Fig. \ref{fig:example}(a) and Fig. \ref{fig:example}(b). In Fig. \ref{fig:example}(c) and Fig. \ref{fig:example}(d), the red boxes and green boxes literally overlap, demonstrating that our proposed method performs well in high SNR regimes.

\subsection{Robustness Analysis}
\begin{figure}[htp]
    \centering
    \includegraphics[width=0.5\textwidth]{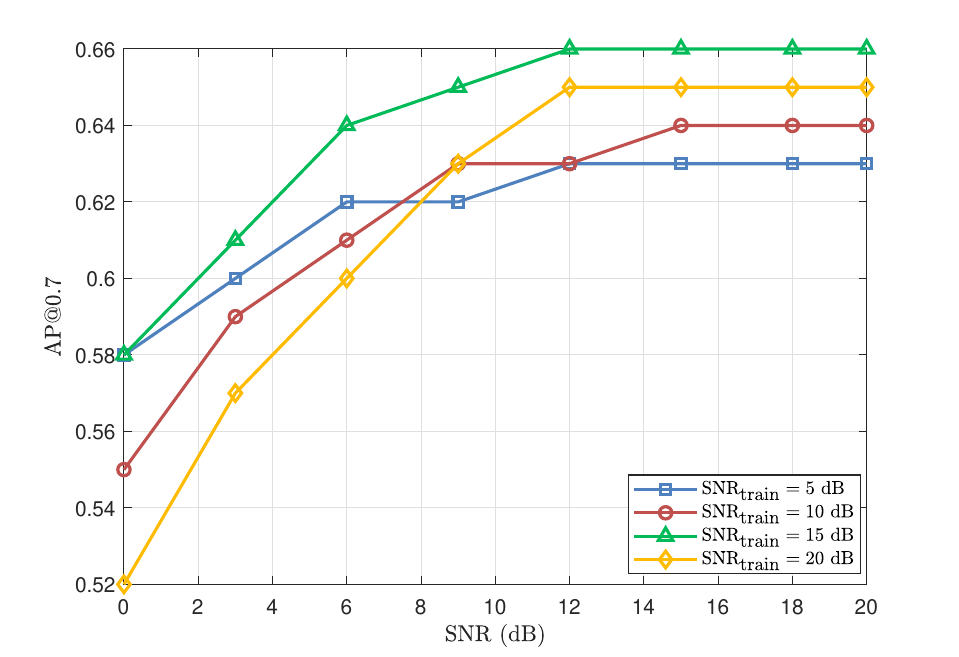}
    \caption{Performance of our proposed method with different training SNRs over an AWGN channel evaluated in terms of AP@0.7.}
    \label{fig:snr_train}
\end{figure}
In this section, we first train our proposed method with different SNRs. Fig. \ref{fig:snr_train} illustrates the performance of our proposed method with different training SNRs over an AWGN channel, which is evaluated in terms of AP$@0.7$. We observe that the training SNRs have a substantial impact on the test performance.  On the one hand, a high training SNR would decrease the robustness of the model to noise, resulting in poor performance in low SNR regimes. On the other hand, a low training SNR can increase robustness but may affect the accuracy of performance in high SNR regimes. For instance, when $\mathrm{SNR_{train}}$ is set as 5 or 10 dB, the performance is not satisfactory at high SNRs, which demonstrates that it is important to choose an appropriate $\mathrm{SNR_{train}}$.
\begin{figure}[htp]
    \centering
    \includegraphics[width=0.5\textwidth]{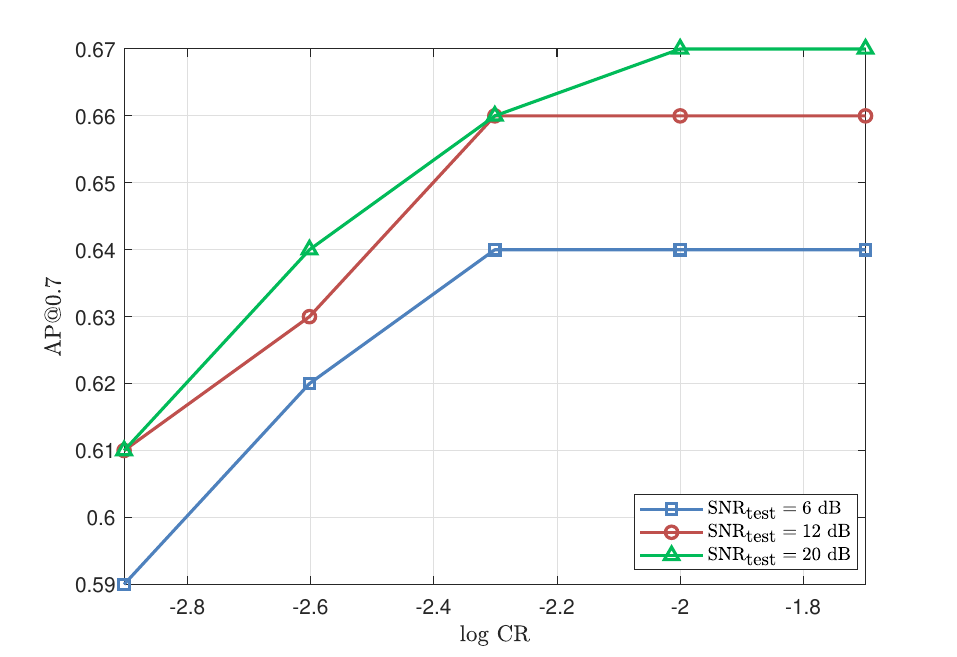}
    \caption{Performance of our proposed method with different compression rates over an AWGN channel evaluated in terms of AP@0.7.}
    \label{fig:cr}
\end{figure}

Fig. \ref{fig:cr} illustrates the performance of our proposed method with different compression rates over an AWGN channel, which is evaluated in terms of AP$@0.7$. We train our model under $\text{CR}=0.005$ and test ours under the condition of different compression rates without extra training. As Fig. \ref{fig:cr} illustrates, despite the mismatched compression rates, our proposed system still maintains excellent performance, which demonstrates its robustness. Meanwhile, since CR represents the compression ratio, which indicates the size of the transmitted data, the overall model exhibits significant performance improvement in both low SNR and high SNR scenarios, as the compression ratio increases. However, with the increase of CR, the performance under different $\mathrm{SNR_{test}}$ would reach the upper bound due to the full exploitation of semantic information. Hence, only a few parts of the feature map are key to the intermediate fusion and the 3D detection, demonstrating that the communication bandwidth can be reduced significantly by our proposed method. 

Moreover, our proposed system, which is trained in only one channel model, is evaluated over different channel models, including TDL-A30, TDL-B100, TDL-C, and TDL-D. On the one hand, the $\text{DS}_{\text{disired}}$ of TDL-A30 and TDL-B100 is 30 ns and 100 ns, which represents short delay spread and nominal delay spread. On the other hand, the $\text{DS}_{\text{disired}}$ of TDL-C and TDL-D is 300 ns, which represents a long delay spread. The PDPs of the four TDL models are normalized to have a total energy of one for simplified evaluations. 

\begin{figure}[htp]
    \centering
    \includegraphics[width=0.5\textwidth]{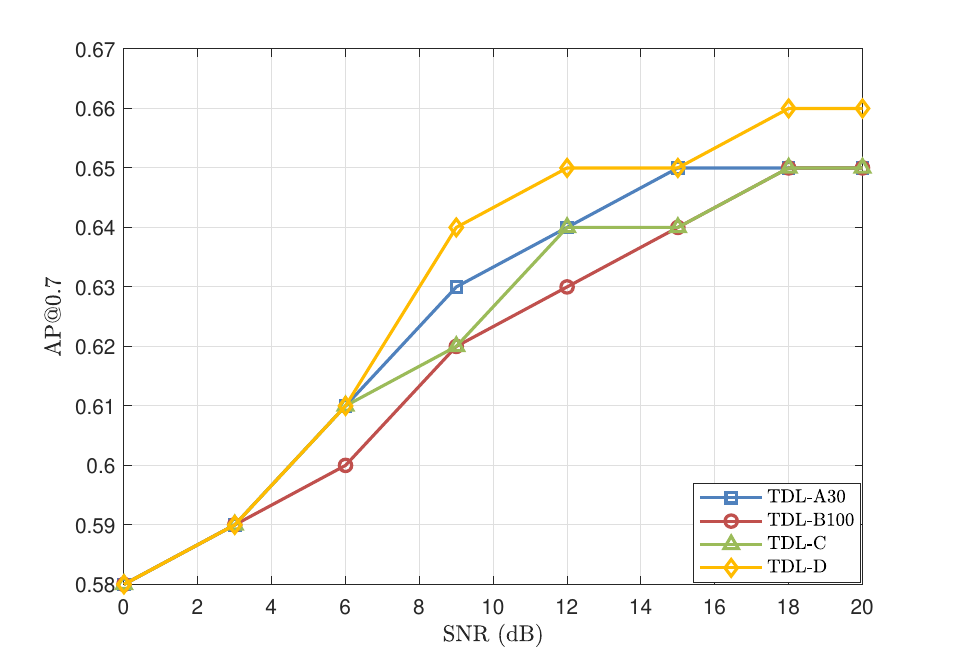}
    \caption{Performance of our proposed method in various channel models.}
    \label{fig:general}
\end{figure}

As Fig. \ref{fig:general} illustrates, despite being trained for a specific channel model, our proposed method is able to learn robust coded representations of the semantic information that are resilient to various channel models. Despite that the delay spreads of four channel model are different from each other, they all reach the upper bound in high SNR regimes. However, due to the difference in the channel models, TDL-D reaches the upper bound AP$@0.7=0.66$ while TDL-B100 and TDL-C reach the upper bound AP$@0.7=0.65$, which has a slight gap. This is because there is a direct path between the transmitter and the receiver in LOS channels (TDL-D), allowing the signal to propagate directly without significant interference or attenuation along the propagation path. Hence, due to the reduced multipath effects in LOS channels, the signal is more stable and reliable since it suffers from less distortion caused by multipath fading. Contrary to the LOS channel, the transmitted signal over the NLOS channels (TDL-A30, TDL-B100, TDL-C) is more sensitive to the distortion caused by the multipath fading and noise, which leads to worse performance. Overall, despite slight performance differences between different channel models, our proposed method is proven to perform reasonably well in all SNR regimes even in the mismatched channel models without any fine-tuning, which demonstrates our generality. Furthermore, our semantic communication system shows its potential to maintain reliable and robust communication without requiring specialized adaptations, contributing to the scalability and versatility of the system.

\section{Conclusion}
We have proposed a novel JSCC architecture for LiDAR point clouds transmission and intermediate fusion over wireless channels. In this architecture, the semantic encoder maps the input LiDAR point clouds directly to channel inputs. The encoder and the decoder functions are modeled as complementary CNNs and trained in an end-to-end manner to minimize the cooperative perception loss and reconstruction loss. Besides AWGN and Rayleigh channel fading channels, we also integrate the JSCC scheme with explicit OFDM blocks to overcome the time-varying multipath fading channel. Our approach involves a meticulous design of the decoder, leveraging expert domain knowledge (e.g., channel estimation and equalization). The simulation results demonstrate the superiority of the proposed model in various channel models, which outperforms the combination of state-of-the-art conventional high-performing channel codes and OFDM systems. Our proposed method demonstrates its generality by learning robust coded representations of semantic information that remains resilient to various channel models even though we train the model using only one specific channel model.


\bibliographystyle{elsarticle-num}

\bibliography{journal_dc.bib}

\end{document}